\documentclass[aip,graphicx]{revtex4-1}

\draft % marks overfull lines with a black rule on the right

\begin{document}

\title{Tetrads in $SU(N)$ Yang-Mills geometrodynamics} %Title of paper

\author{Alcides Garat}
%\email[]{Your e-mail address}
%\homepage[]{Your web page}
%\thanks{}
%\altaffiliation{}
\affiliation{1. Former Professor at Universidad de la Rep\'{u}blica, Instituto de F\'{\i}sica, Facultad de Ingenier\'{\i}a, J. Herrera y Reissig 565, 11300 Montevideo, Uruguay.}
\email[]{garat.alcides@gmail.com}
\date{\today}
%Universidad de la Rep\'{u}blica, Instituto de F\'{\i}sica, Facultad de Ingenier\'{\i}a, J. Herrera y Reissig 565, 11300 Montevideo, Uruguay. \footnote {Former Professor at Universidad de la Rep\'{u}blica, Instituto de F\'{\i}sica, Facultad de Ingenier\'{\i}a, J. Herrera y Reissig 565, 11300 Montevideo, Uruguay.}
\begin{abstract}
The discovery of the SU(3) symmetry was fundamental as to establishing an ordering principle in particle physics. We already studied how to couple the SU(3) symmetry to the gravitational field in four-dimensional curved Lorentzian spacetimes. The multiplets of equal quantum numbers are translated through natural elements in Riemannian geometry into local multiplets of equal gravitational field. As quark physics developed since the seventies, it was necessary to incorporate new symmetries to the models, that ensued in the incorporation of new quantum numbers like Charm, for example. Charm is an additive quantum number like isospin $T_{3}$ and hypercharge $Y$ and the standard $T_{3}-Y$ diagrams were extended onto another third axis. Then, instead of the fundamental triplet we have a quartet $\{u,d,s,c\}$ as the smallest representation of the symmetry group, leading to the introduction of SU(4) as the new group of symmetries. In this paper we will not restrict ourselves exclusively to the symmetry group SU(4) and we will set out to analyze the coupling of the $SU(N)$ symmetry to the gravitational field. To this end new tetrads will be introduced as we did for the $SU(3) \times SU(2) \times U(1)$ case. These tetrads have outstanding properties that enable these constructions. New theorems will be proved regarding the isomorphic nature of these local symmetry gauge groups and tensor products of groups of local tetrad transformations. This is a paper about grand field unification in four-dimensional curved Lorentzian spacetimes.
\end{abstract}

\pacs{}% insert suggested PACS numbers in braces on next line

\maketitle %\maketitle must follow title, authors, abstract and \pacs

\section{Introduction}
\label{intro}

The advent of the eightfold way and the quark model brought forward the consideration of $SU(N)$ Lie groups and their representations. Because of the development of Quantum Field Theories and the experimental success of its predictions including many new symmetries. Many new particles associated to $SU(N)$ representations were found and that is how these groups and their representations came into play in fundamental physics. This process began by the beginning of the 1930s  when the isospin SU(2) model was first considered. The myriad of experiments involving electrodynamics, weak interactions and chromodynamics brought about the necessity to include the symmetries associated to Abelian and non-Abelian gauge groups. There is an enormous amount of literature on these subjects. We just cite a limited number of books and papers that themselves include ample sets of references \cite{DG}$^{-}$\cite{MC}. There was a parallel development of the theory of General Relativity associated to gravitational phenomena in the large scale. General Relativity is another successful theory that explains a myriad of interactions involving gravitational fields like Black Holes, Gravitational Lensing, Collisions of Neutron Stars, Gamma Ray Bursts, and also developing theories like Quantum Gravity, etc. There is another enormous library of literature on this subject, we just cite a few books and papers which themselves include ample sets of references \cite{PA1}$^{-}$\cite{HS}. It is very well known the difficulty in creating a single unified framework encompassing General Relativity and Quantum Field Theory. Therefore, we set out to find first in a classical setup and then incorporating a quantum context how we can realize all the local spacetime symmetries from General Relativity and the local gauge symmetries from Quantum Field Theories in a four-dimensional curved Lorentzian spacetime with signature $(-+++)$. The natural structures to start studying this intricate problem are the Einstein-Maxwell spacetimes with a set of classical differential equations given by,

\begin{eqnarray}
f^{\mu\nu}_{\:\:\:\:\:;\nu} &=& 0 \label{EM1}\\
\ast f^{\mu\nu}_{\:\:\:\:\:;\nu} &=& 0 \label{EM2}\\
R_{\mu\nu} &=& f_{\mu\lambda}\:\:f_{\nu}^{\:\:\:\lambda}
+ \ast f_{\mu\lambda}\:\ast f_{\nu}^{\:\:\:\lambda}\ , \label{EM3}
\end{eqnarray}

If $F_{\mu\nu}$ is the electromagnetic field then $f_{\mu\nu}= (G^{1/2} / c^2) \: F_{\mu\nu}$ is the geometrized
electromagnetic field, where $\ast f_{\mu\nu}={1 \over 2}\:\epsilon_{\mu\nu\sigma\tau}\:f^{\sigma\tau}$ is the dual tensor of $f_{\mu\nu}$, see section IX in reference \cite{A}. The symbol $``;''$ stands for covariant derivative with respect to the metric tensor $g_{\mu\nu}$. It is a matter of simple algebra to prove a system of results that will have profound further consequences in grand unification and that were overlooked for the past century. The Einstein-Maxwell stress-energy tensor according to equation (14a) in reference \cite{MW}, can be written as,

\begin{equation}
T_{\mu\nu}= f_{\mu\lambda}\:\:f_{\nu}^{\:\:\:\lambda} + \ast f_{\mu\lambda}\:\ast f_{\nu}^{\:\:\:\lambda}\ ,\label{TEM}
\end{equation}

The alternating tensor $\epsilon_{\mu\nu\sigma\tau}$ is studied in appendix IX of reference \cite{A}. The local duality rotation given by equation (59) in\cite{MW},

\begin{equation}
f_{\mu\nu} = \xi_{\mu\nu} \: \cos\alpha + \ast\xi_{\mu\nu} \: \sin\alpha\ ,\label{dr}
\end{equation}

transforms the stress-energy tensor expression (\ref{TEM}) in terms of the extremal field,

\begin{equation}
T_{\mu\nu}=\xi_{\mu\lambda}\:\:\xi_{\nu}^{\:\:\:\lambda} + \ast \xi_{\mu\lambda}\:\ast \xi_{\nu}^{\:\:\:\lambda}\ .\label{TEMDR}
\end{equation}

The extremal field $\xi_{\mu\nu}$ and the local scalar complexion $\alpha$ are defined through equations (22-25) in \cite{MW}.
In terms of these variables framework it is possible to find a tetrad in which the stress-energy tensor is diagonal. This tetrad would simplify the analysis of the geometrical properties of the electromagnetic field. There are four tetrad vectors that at every point in spacetime diagonalize the stress-energy tensor (\ref{TEMDR}) in geometrodynamics,

\begin{eqnarray}
V_{(1)}^{\alpha} &=& \xi^{\alpha\lambda}\:\xi_{\rho\lambda}\:X^{\rho}
\label{V1}\\
V_{(2)}^{\alpha} &=& \sqrt{-Q/2} \: \xi^{\alpha\lambda} \: X_{\lambda}
\label{V2}\\
V_{(3)}^{\alpha} &=& \sqrt{-Q/2} \: \ast \xi^{\alpha\lambda} \: Y_{\lambda}
\label{V3}\\
V_{(4)}^{\alpha} &=& \ast \xi^{\alpha\lambda}\: \ast \xi_{\rho\lambda}
\:Y^{\rho}\ ,\label{V4}
\end{eqnarray}

where $Q=\xi_{\mu\nu}\:\xi^{\mu\nu}=-\sqrt{T_{\mu\nu}T^{\mu\nu}}$ according to equations (39) in \cite{MW}. $Q$ is assumed not to be zero,
because we are dealing with non-null electromagnetic fields. Non-null we clarify means basically that $f_{\mu\nu}\:f^{\mu\nu}\neq0$ and $\ast f_{\mu\nu}\:f^{\mu\nu}\neq0$. In turn and by definitions these last equations imply that $\xi_{\mu\nu}\:\xi^{\mu\nu}\neq0$. This is the most general set of four vector fields that diagonalizes the stress-energy tensor in geometrodynamics. Let us introduce some names. The tetrad vectors have two essential components. For instance in vector $V_{(1)}^{\alpha}$ there are two main structures. First, the skeleton, in this case $\xi^{\alpha\lambda}\:\xi_{\rho\lambda}$, and second, the gauge vector $X^{\rho}$. The gauge vectors it was proved in manuscript \cite{A} could be anything that does not make the tetrad vectors trivial. That is, the tetrad (\ref{V1}-\ref{V4}) diagonalizes the stress-energy tensor for any non-trivial gauge vectors $X^{\mu}$ and $Y^{\mu}$. It was therefore proved that we can make different choices for $X^{\mu}$ and $Y^{\mu}$. We are free to choose the vector fields $X^{\alpha}$ and $Y^{\alpha}$, as long as the four vector fields (\ref{V1}-\ref{V4}) are not trivial. Extremal fields are essentially electric fields and they satisfy,

\begin{equation}
\xi_{\mu\nu} \ast \xi^{\mu\nu}= 0\ . \label{i0}
\end{equation}

Equation (\ref{i0}) is a condition imposed on the extremal field $\xi_{\mu\nu} = e^{-\ast \alpha} f_{\mu\nu}\ = \cos\alpha\:f_{\mu\nu} - \sin\alpha\:\ast f_{\mu\nu}$ which is the inverse of equation (\ref{dr}) and then the explicit expression for the complexion is found, given by $\tan(2\alpha) = - f_{\mu\nu}\:\ast f^{\mu\nu} / f_{\lambda\rho}\:f^{\lambda\rho}$. As antisymmetric fields in a four dimensional Lorentzian spacetime, the extremal fields also verify the identity,

\begin{eqnarray}
\xi_{\mu\alpha}\:\xi^{\nu\alpha} -
\ast \xi_{\mu\alpha}\: \ast \xi^{\nu\alpha} &=& \frac{1}{2}
\: \delta_{\mu}^{\:\:\:\nu}\ Q \ .\label{i1}
\end{eqnarray}

It can be proved that condition (\ref{i0}) and through the use of the general identity,

\begin{eqnarray}
A_{\mu\alpha}\:B^{\nu\alpha} -
\ast B_{\mu\alpha}\: \ast A^{\nu\alpha} &=& \frac{1}{2}
\: \delta_{\mu}^{\:\:\:\nu}\: A_{\alpha\beta}\:B^{\alpha\beta}  \ ,\label{ig}
\end{eqnarray}

which is valid for every pair of antisymmetric tensors in a four-dimensional Lorentzian spacetime \cite{MW}, when applied to the case $A_{\mu\alpha} = \xi_{\mu\alpha}$ and $B^{\nu\alpha} = \ast \xi^{\nu\alpha}$ yields the equivalent condition,

\begin{eqnarray}
\xi_{\alpha\mu}\:\ast \xi^{\mu\nu} &=& 0\ ,\label{i2}
\end{eqnarray}

which is equation (64) in \cite{MW}. It is evident that identity (\ref{i1}) is a special case of (\ref{ig}). Two identities in the extremal field will be used extensively in this work, in particular, to prove that tetrad (\ref{V1}-\ref{V4}) diagonalizes the stress-energy tensor. Using equations (\ref{i0}-\ref{i2}) it is straightforward to prove that the four vectors (\ref{V1}-\ref{V4}) are orthogonal. When we make iterative use of (\ref{i1}) and (\ref{i2}) we find,

\begin{eqnarray}
V_{(1)}^{\alpha}\:T_{\alpha}^{\:\:\:\beta} &=& \frac{Q}{2}\:V_{(1)}^{\beta}
\label{EV1}\\
V_{(2)}^{\alpha}\:T_{\alpha}^{\:\:\:\beta} &=& \frac{Q}{2}\:V_{(2)}^{\beta}
\label{EV2}\\
V_{(3)}^{\alpha}\:T_{\alpha}^{\:\:\:\beta} &=& -\frac{Q}{2}\:V_{(3)}^{\beta}
\label{EV3}\\
V_{(4)}^{\alpha}\:T_{\alpha}^{\:\:\:\beta} &=& -\frac{Q}{2}\:V_{(4)}^{\beta}\ .
\label{EV4}
\end{eqnarray}

In \cite{MW} the stress-energy tensor was diagonalized through the use of a Minkowskian frame in which the expression for this tensor was given in equations (34) and (38). In this work, we give the explicit expression for the tetrad in which the stress-energy tensor is diagonal in a local and covariant way. The freedom we have to choose the vector fields $X^{\alpha}$ and $Y^{\alpha}$, represents available freedom that we have to choose the tetrad. If we make use of equations (\ref{i1}) and (\ref{i2}), it is straightforward to prove that (\ref{V1}-\ref{V4}) is a set of orthogonal vectors. The freedom to choose the vector fields $X^{\alpha}$ and $Y^{\alpha}$ is a gauge freedom and it was proved through a set of theorems in reference \cite{A}, that these freedom is equivalent to the freedom to gauge the electromagnetic potentials in electrodynamics. In the Einstein-Maxwell context we have available two vector potentials which are not independent from each other but that exist in the end, as demonstrated in manuscript \cite{CF}. In geometrodynamics, the Maxwell equations, $f^{\mu\nu}_{\:\:\:\:\:;\nu} = 0$ and $\ast f^{\mu\nu}_{\:\:\:\:\:;\nu} = 0$ are telling us that two potential vector fields exist \cite{CF}, $f_{\mu\nu} = A_{\nu ;\mu} - A_{\mu ;\nu}$ and
$\ast f_{\mu\nu} = \ast A_{\nu ;\mu} - \ast A_{\mu ;\nu}$. For instance, in the Reissner-Nordstr\"{o}m geometry the only non-zero electromagnetic tensor component is $f_{tr}=A_{r;t} - A_{t;r}$ and its dual $\ast f_{\theta\phi}=\ast A_{\phi;\theta} - \ast A_{\theta;\phi}$. The symbol $``;''$ stands for covariant derivative with respect to the metric tensor $g_{\mu\nu}$ and the star in $\ast A_{\nu}$ is just a name, not the dual operator, meaning that $\ast A_{\nu ;\mu} = (\ast A_{\nu})_{;\mu}$. The vector fields $A^{\alpha}$ and $\ast A^{\alpha}$ represent a possible choice in geometrodynamics for the vectors $X^{\alpha}$ and $Y^{\alpha}$. It is not meant that the two vector fields have independence from each other, it is just a convenient choice for a particular example. A further justification for the choice $X^{\alpha}=A^{\alpha}$ and $Y^{\alpha}=\ast A^{\alpha}$ could be illustrated through the Reissner-Nordstr\"{o}m geometry. In this particular geometry, $f_{tr}=\xi_{tr}$ and $\ast f_{\theta\phi}=\ast \xi_{\theta\phi}$, therefore, $A_{\theta}=0$ and $A_{\phi}=0$. Then, for the last two tetrad vectors (\ref{V3}-\ref{V4}), the choice $Y^{\alpha}=\ast A^{\alpha}$ becomes meaningful under the light of this particular extreme case, when basically there is no magnetic field. Once we make the choice $X^{\alpha}=A^{\alpha}$ and $Y^{\alpha}=\ast A^{\alpha}$ the question about the geometrical implications of electromagnetic gauge transformations of the tetrad vectors (\ref{V1}-\ref{V4}) arises. We first notice that a local electromagnetic gauge transformation of the ``gauge vectors'' $X^{\alpha}=A^{\alpha}$ and $Y^{\alpha}=\ast A^{\alpha}$ can be just interpreted as a new choice for the gauge vectors $X_{\alpha} = A_{\alpha} + \Lambda_{,\alpha}$ and $Y_{\alpha} = \ast A_{\alpha} + \ast \Lambda_{,\alpha}$. When we make the transformation, $A_{\alpha} \rightarrow A_{\alpha} + \Lambda_{,\alpha}$, $f_{\mu\nu}$ remains invariant, and the transformation, $\ast A_{\alpha} \rightarrow \ast A_{\alpha} + \ast \Lambda_{,\alpha}$, leaves $\ast f_{\mu\nu}$ invariant, as long as the functions $\Lambda$ and $\ast \Lambda$ are scalars. It is valid to ask how the tetrad vectors (\ref{V1}-\ref{V2}) will transform under $A_{\alpha} \rightarrow A_{\alpha} + \Lambda_{,\alpha}$ and (\ref{V3}-\ref{V4}) under $\ast A_{\alpha} \rightarrow \ast A_{\alpha} + \ast \Lambda_{,\alpha}$. Schouten defined what he called, a two-bladed structure in a spacetime \cite{SCH}. These local orthogonal blades or planes are the planes determined by the pairs ($V_{(1)}^{\alpha}, V_{(2)}^{\alpha}$) for the plane or blade one and ($V_{(3)}^{\alpha}, V_{(4)}^{\alpha}$) for the plane or blade two. On reference \cite{A} theorems were proved stating that the local group of electromagnetic gauge transformations is isomorphic to the local group of tetrad transformations LB1 on the local blade one and independently to the local group LB2 of tetrad transformations on the local blade two. LB1 as a group has two sheets. The first sheet is LB1 proper made up of the boosts SO(1,1) plus a discrete full inversion or minus the two by two identity. LB1 improper is the second sheet made up of LB1 proper composed with a non-Lorentzian reflection that we called the switch or flip. This flip is a discrete transformation given by $\Lambda^{o}_{\:\:o} = 0$, $\Lambda^{o}_{\:\:1} = 1$, $\Lambda^{1}_{\:\:o} = 1$, $\Lambda^{1}_{\:\:1} = 0$. We notice that this discrete transformation is not a Lorentz transformation because it is a reflection. LB2 is SO(2). The local group algebra of electromagnetic gauge transformations is isomorphic to the new groups LB1 and LB2, independently. We know that the local algebra of the U(1) group, that is the real numbers $\mathcal{R}$ or local scalars $\Lambda$ is homomorphic to the local group of Abelian electromagnetic gauge transformations U(1). LB1 is the group of local tetrad transformations comprised by SO(1,1) plus two different kinds of discrete transformations. One of these two discrete transformations is the full inversion or minus the identity two by two. All the elements that arise by composing the full inversion and SO(1,1) are present in the image of this mapping even though the full inversion itself is not, however, it is an accumulation point in the image of the mapping. The other discrete transformation is not Lorentzian because it is a reflection or flip with zeroes in the diagonal and ones off-diagonal also two by two. The local group algebra of electromagnetic gauge transformations is isomorphic to the local group of tetrad transformations LB2 on the orthogonal local plane as well. LB2 is SO(2) minus the full inversion. The full inversion is however an accumulation point in the image of this mapping. The existence of these isomorphisms between the local algebra and these LB1 and LB2 groups on local orthogonal planes is possible since in reality LB2 is SO(2) minus the full inversion or minus the identity two by two and similar for LB1. It was also proved by transitivity that LB1 is a double covering of LB2. There is a homomorphism between LB1 and LB2. By topological closure we can map the missing full inversion in LB1 to the missing full inversion in LB2 since both are simultaneous limits of one to one sequences. This alone is a discovery in group theory \cite{RGLG,RG,JS,NC} and fundamental physics. It has even further implications that go directly to the core of gauge theories and quantum theories. Because it contradicts in a manifest and direct way the results of the no-go theorems of the sixties \cite{SWNG,LORNG,CMNG}. For essentially two reasons. We read from reference \cite{CMNG} ``S (the scattering matrix) is said to be Lorentz-invariant if it possesses a symmetry group locally isomorphic to the Poincar\`{e} group P.\ldots A symmetry transformation is said to be an internal symmetry transformation if it commutes with P. This implies that it acts only on particle-type indices, and has no matrix elements between particles of different four-momentum or different spin. A group composed of such transformations is called an internal symmetry group''. The local electromagnetic gauge group algebra of transformations U(1) has been proven to be isomorphic to local groups of tetrad transformations on both the orthogonal planes one and two. We remind ourselves that the local orthogonal planes one and two are spanned by the tetrad vectors (\ref{V1}-\ref{V2}) and (\ref{V3}-\ref{V4}) respectively. All the vectors on these local planes are eigenvectors of the Einstein-Maxwell stress-energy tensor (\ref{TEMDR}). These local Lorentz transformations LB1 proper and SO(2) are Lorentz transformations and even though LB1 improper is not composed of Lorentz transformations, it is composed of spacetime transformations since it is LB1 proper composed with the switch or flip which is a discrete reflection, see reference \cite{A}. Therefore the local Lorentz group of spacetime transformations cannot commute with LB1 or LB2 since Lorentz transformations on a local plane do not commute with Lorentz transformations on another local plane. In addition it has been proven by transitivity that $LB1 = \{LB1 proper, LB1 improper\}$ is a double covering of SO(2). There is a two to one homomorphism between LB1 and $LB2 = SO(2)$. The conclusion is that the no-go theorems \cite{SWNG,LORNG,CMNG} are incorrect since from the outset these theorems make incorrect assumptions or hypotheses. In manuscripts \cite{A,A2,A3,ASU3} we extended these Abelian results to the non-Abelian cases $SU(2) \times U(1)$ and $SU(3) \times SU(2) \times U(1)$. In this paper we will expand these results to the $SU(N)$ case. We want to develop suitable tools in order to understand the nature of gravitational fields in the presence of local Abelian and non-Abelian Yang-Mills fields. In addition we find fundamental results in group theory and new techniques in dealing with gauge invariant diagonalization of stress-energy tensors \cite{A}$^{-}$\cite{MW}. In the non-Abelian case through a new kind of duality transformations. We also establish explicitly the relationship between gauge fields and gravity through non-trivial new tetrads. Differential equations will also be simplified.  Other future applications are also possible. For instance, the study of the kinematics in these spacetimes \cite{CBDW}$^{-}$\cite{RGE} that is, the search for a ``connection'' to the theory of embeddings, time slicings \cite{KK}$^{-}$\cite{FAEP}, the initial value formulation \cite{JWY1}$^{-}$\cite{CMDWYCB}, the Cauchy evolution \cite{AEO}$^{-}$\cite{LSJWY}, etc. This manuscript is organized as follows.

\section{Quotient Space}
\label{QS}

In the case under study, that is local $SU(N)$, we cannot proceed as in the $U(1)$ or $SU(2)$ case. Analogous to the $SU(3)$ case we have to consider $N \times N$ matrices and we have to develop a different strategy. We will use the notion of quotient space, in particular the relation \cite{MN}$^{,}$\cite{CI},

\begin{equation}
SU(N) / SU(N-1) \cong S^{2N-1}\ . \label{quotiantu}
\end{equation}

Let us first proceed to understand what we will do through a simple example,

\begin{equation}
SO(3) / SO(2) \cong S^{2}\ . \label{quotianto}
\end{equation}

We can generate all $SO(3)$ transformations by fixing a direction in $S^{2}$, that is, choosing a unit vector in one special direction, and then performing all possible $SO(2)$ transformations in the orthogonal plane. If we repeat this process for all directions in $S^{2}$, then we would be spanning the whole $SO(3)$ group of local transformations. We can use this notion in order to implement a similar idea for the $SU(N)$ case as we did for the $SU(3)$ case in manuscript \cite{ASU3} which is not possible to visualize like the $SO(3)$ procedure.
%Then, we proceed to choose a unit vector in $S^{5}$. Let us say that $(\cos\gamma_{4},\cos\gamma_{5},\cos\gamma_{6},\cos\gamma_{7},\cos\gamma_{8})$ is a local vector in $S^{5}$ such that $\sum_{i=4}^{8} \cos^{2}\gamma_{i} = 1$. All the $\cos\gamma_{i},\:i=4\cdots8$ are local scalars.
Let us choose as our analog to $SO(2)$, the $SU(N-1)$ subalgebra to the $SU(N)$ algebra generated by the standard generators as given in reference \cite{GMS} chapter XI. Let us call these $(N-1)^{2}-1$ generators $X_{1},\ldots, X_{(N-1)^{2}-1}$. They obey the commutation relations from chapter XI in reference \cite{GMS} $[X_{i},X_{j}] = 2\imath\:f_{ijk}\:X_{k}$, where the sum convention was applied on $k$. Then it is clear that we can call

\begin{eqnarray}
S_{sub} = \exp \{ (\imath/2)\:\sum_{i=1}^{(N-1)^{2}-1} \theta_{i}\:X_{i} \} \ , \label{SU(N-1)}
\end{eqnarray}

where $\theta_{i},\: i=1 \ldots (N-1)^{2}-1$ are local scalars. In (\ref{SU(N-1)}), the matrices $S_{sub}$ are local. Next, we build the local $SU(N)$ group element object

\begin{eqnarray}
S_{N} = \exp \{ (\imath/2)\:\sum_{i=(N-1)^{2}}^{N^{2}-1} \psi_{i}\:X_{i} \} \ , \label{SU(N)}
\end{eqnarray}

where the $\psi_{i},\: i=(N-1)^{2} \ldots N^{2}-1$ are all local scalars. The $X_{i}, i=(N-1)^{2} \ldots N^{2}-1$ are the remaining $2N-1$, $SU(N)$ group generators. The first $(N-1)^{2}-1$ were associated to the $SU(N-1)$ subalgebra. We call $S_{N}$ the $SU(N)$ local group coset elements that represent a direction in the $S^{2N-1}$ sphere about which $SU(N-1)$ ``rotations'' are performed, see sections \ref{appendixI} and \ref{appendixII} for details. We remind ourselves that for every choice of a vector in $S^{2N-1}$ we perform all possible local gauge $SU(N-1)$ transformations. We repeat this process for every vector in $S^{2N-1}$. Now, every possible direction in the $S^{2N-1}$ sphere is represented by a different local $SU(N)$ coset element $S_{N}$. They span a space, a manifold, not a group by themselves. In the end we translate every $SU(N)$ local gauge transformation into a product of transformations, following the ideas of Cartan, see chapter VII in reference \cite{RGLG}. In our notation every local $SU(N)$ group element will be written as $S_{N}\:S_{sub}$. This way we will be able to ``link'' the local $SU(N)$ to the local $SU(N-1)$ and establish local group isomorphisms between the local $SU(N)$ and the local tensor products of LB1 or LB2. Because if we can express an $SU(N)$ local group element as a product like $S_{N}\:S_{sub}$, a local coset element times a local subgroup element, then we can repeat this process inductively until we reach $SU(2)$ like $S_{sub} = S_{N-1}\:\ldots \:S_{3}\:S_{2}$ at which point we can apply all the results of manuscript \cite{ASU3}. That is the idea underlying our theorem proofs in the next section \ref{TSUN}.

\section{Tetrads in $SU(N)$}
\label{TSUN}

We proceed to introduce the system of equations that represents the starting point of this tetrad construction.

\begin{eqnarray}
R_{\mu\nu} &=& T^{(ymsuN)}_{\mu\nu} + \ldots + T^{(ymsu3)}_{\mu\nu} + T^{(ymsu2)}_{\mu\nu} + T^{(em)}_{\mu\nu}\label{eyme}\\
f^{\mu\nu}_{u(1)\:\:\:\:;\nu} &=& 0 \label{EM1}\\
\ast f^{\mu\nu}_{u(1)\:\:\:\:;\nu} &=& 0 \label{EM2}\\
f^{k_{2}\mu\nu}_{su(2)\:\:\:\:\:\:\:\mid \nu} &=& 0 \label{ymsu21}\\
\ast f^{k_{2}\mu\nu}_{su(2)\:\:\:\:\:\:\:\mid \nu} &=& 0 \label{ymsu22}\\
f^{k_{3}\mu\nu}_{su(3)\:\:\:\:\:\:\:\mid \nu} &=& 0 \label{ymsu31}\\
\ast f^{k_{3}\mu\nu}_{su(3)\:\:\:\:\:\:\:\mid \nu} &=& 0 \label{ymsu32} \\
&\vdots& \nonumber \\
f^{k_{N}\mu\nu}_{su(N)\:\:\:\:\:\:\:\mid \nu} &=& 0 \label{ymsuN1}\\
\ast f^{k_{N}\mu\nu}_{su(N)\:\:\:\:\:\:\:\mid \nu} &=& 0 \ . \label{ymsuN2}
\end{eqnarray}

where the internal index $k_{2}$ is an $SU(2)$ index running from $k_{2}=1\cdots3$, while $k_{3}$ is an $SU(3)$ index running from $k_{3}=1\cdots8$ and $k_{N}$ is an $SU(N)$ index running from $k_{N}=1\cdots(N^{2}-1)$.  The symbol ``;'' stands for the usual covariant derivative associated with the metric tensor $g_{\mu\nu}$, while $\mid$ stands for gauge covariant derivative. The tensors to the right of equation (\ref{eyme}) are the $SU(N)$,...,$SU(3)$, $SU(2)$ and $U(1)$ standard stress-energy tensors \cite{MC}. First of all let us say that it is clear by now that we can proceed to build local gauge invariant tetrad skeletons using local gauge invariant extremal antisymmetric second rank tensors $\Omega^{\mu\lambda}$ for example, following a similar procedure for $SU(N)$ as in section ``Extremal field in $SU(2)$ geometrodynamics'' in paper \cite{A2} for $SU(2)$. It would be redundant to repeat it here. We will just introduce the new tetrad that we build following the constructions in papers \cite{A}$^{,}$\cite{A2}$^{,}$\cite{A3}$^{,}$\cite{ASU3},

\begin{eqnarray}
Q_{(1)}^{\mu} &=& \Omega^{\mu\lambda}\:\Omega_{\rho\lambda}\:X^{\rho}
\label{S1}\\
Q_{(2)}^{\mu} &=& \sqrt{-Q_{ym}/2} \: \Omega^{\mu\lambda} \: X_{\lambda}
\label{S2}\\
Q_{(3)}^{\mu} &=& \sqrt{-Q_{ym}/2} \: \ast \Omega^{\mu\lambda} \: Y_{\lambda}
\label{S3}\\
Q_{(4)}^{\mu} &=& \ast \Omega^{\mu\lambda}\: \ast\Omega_{\rho\lambda}
\:Y^{\rho}\ ,\label{S4}
\end{eqnarray}

where $Q_{ym} = \Omega_{\mu\nu}\:\Omega^{\mu\nu}$ is considered not to be zero and $\Omega_{\mu\nu}\:\ast \Omega^{\mu\rho} = 0$. Then, let us define the ``gauge'' vector for our $SU(N)$ case,

\begin{equation}
X^{\sigma} = Y^{\sigma} = Tr[\Sigma^{\alpha\beta}\:S_{\alpha}^{\:\:\rho}\: S_{\beta}^{\:\:\lambda}\:\ast \epsilon_{\rho}^{\:\:\sigma}\:\ast \epsilon_{\lambda\tau}\:A^{\tau}] \ .\label{gaugev}
\end{equation}

$\Sigma^{\alpha\beta}$ is the antisymmetric object defined exactly and analogously as in reference \cite{A2} for the $X_{1}, X_{2}, X_{3}$ generators of the $SU(2)$ subalgebra which in this case are $N \times N$ matrices in this present work (see appendix II in reference \cite{A2}). $\Sigma^{\alpha\beta}$ are essentially built with the Pauli matrices. $S_{\alpha}^{\:\:\rho}$ are the local $SU(N-1)$ tetrads defined exactly as in \cite{A2,A3,ASU3} for a $SU(2)$ and $SU(3)$ gauge vectors. Let us remember that $A^{\tau}$ in (\ref{gaugev}) is a $SU(N)$ gauge vector, a $N \times N$ matrix. The structure $S_{\alpha}^{\:\:[\rho}\: S_{\beta}^{\:\:\lambda]}\:\ast \epsilon_{\rho}^{\:\:\sigma}\:\ast \epsilon_{\lambda\tau}$ is invariant under $SU(N-1)$ local gauge transformations. Essentially, because of the $SU(N-1)$ extremal field property \cite{A}$^{,}$\cite{MW}$^{,}$\cite{A2}, $\epsilon_{\mu\sigma}\:\ast \epsilon^{\mu\tau} = 0$. Leaving thus in the contraction with $\ast \epsilon_{\rho\sigma}\:\ast \epsilon_{\lambda\tau}$, and because of property $\epsilon_{\mu\sigma}\:\ast \epsilon^{\mu\tau} = 0$, only the antisymmetric object $S_{2}^{\:\:[\rho}\:S_{3}^{\:\:\lambda]}$, which is locally $SU(N-1)$ gauge invariant by construction and by induction. Let us remember that the object $\Sigma^{\alpha\beta}$ is antisymmetric and contracted with the $SU(N-1)$ tetrads as $\Sigma^{\alpha\beta}\:S_{\alpha}^{\:\:\rho}\: S_{\beta}^{\:\:\lambda}$ inside the local gauge vector (\ref{gaugev}). We already knew when we studied the $SU(3)$ case (see the second appendix in \cite{A2}) that the tetrads $S_{\beta}^{\:\:\lambda}$ were themselves $SU(2)$ tetrads invariant under local $U(1)$ gauge transformations \cite{A2}. We mean by this last remark, their tetrad skeletons and specially defined gauge vectors. Therefore, the $SU(3)$ gauge vectors $X^{\sigma} = Y^{\sigma}$ are locally invariant under $SU(2) \times U(1)$ local gauge transformations. In our present case the $SU(N)$ gauge vectors $X^{\sigma} = Y^{\sigma}$ are locally invariant under $SU(N-1) \times \ldots \times SU(2) \times U(1)$ local gauge transformations. This is fundamental since it enables us to introduce local $SU(N)$ tetrad gauge transformations independently of $SU(N-1) \times \ldots \times SU(2) \times U(1)$ local gauge transformations and talk about $SU(N) \times SU(N-1) \times \ldots \times SU(3) \times SU(2) \times U(1)$ Yang-Mills geometrodynamics. It is important not to get confused by the product of exponentials in $SU(N)$ where one of the factors is a $SU(N-1)$ element of the subalgebra on one hand, and ``pure'' $SU(N-1)$ gauge transformations on the other hand. The $SU(N-1)$ subalgebra ``operates'' through the gauge vector (\ref{gaugev}), while the ``pure'' $SU(N-1)$ gauge transformations get into play through a ``pure'' $SU(N-1)$ gauge vector. When these two different gauge vectors for $SU(N-1)$ and $SU(N)$ tetrads are added as a possible choice of gauge vector in order to gauge the tetrad vectors, their inherent transformations are independent and the transformations LB1 and LB2 they induce commute between themselves, see Appendix III in reference \cite{A2}. Next we proceed to study the tetrad gauge transformations of the gauge vector (\ref{gaugev}). It is straightforward to notice that we can borrow all the analysis previously done in the section gauge geometry in reference \cite{A2}. Nonetheless it is important to pay attention to the nature of the gauge transformed local gauge vector (\ref{gaugev}) under the sequence of local $SU(N)$ transformations introduced in section \ref{QS}. This way, we are able to study the local gauge transformation of gauge vector (\ref{gaugev}) under any local $SU(N)$ gauge transformation since the product of exponentials covers all of the local $SU(N)$. Then, we proceed to transform (\ref{gaugev}) in a sequence, first with $S_{sub}$, the local subalgebra group element, and then with $S_{N}$, the local coset representative,

\begin{eqnarray}
Tr[S_{N}\:S_{sub}\:\Sigma^{\alpha\beta}\:S_{sub}^{-1}\:S_{N}^{-1}\:S_{\alpha}^{\:\:\rho}\: S_{\beta}^{\:\:\lambda}\:\ast \epsilon_{\rho\sigma}\:\ast \epsilon_{\lambda\tau}\:A^{\tau}] + \nonumber \\ {\imath \over g} \:Tr[S_{N}\:S_{sub}\:\Sigma^{\alpha\beta}\:S_{sub}^{-1}\:S_{N}^{-1}\:S_{\alpha}^{\:\:\rho}\: S_{\beta}^{\:\:\lambda}\:\ast \epsilon_{\rho\sigma}\:\ast \epsilon_{\lambda\tau}\:\partial^{\tau}\:(S_{N}\:S_{sub})\:(S_{N}\:S_{sub})^{-1}]
\label{S1R}
\end{eqnarray}

\begin{eqnarray}
\exp \{ (\imath/2)\:\sum_{i=1}^{N^{2}-1} x_{i}\:X_{i} \}  = \exp \{ (\imath/2)\:\sum_{j=(N-1)^{2}}^{N^{2}-1} y_{j}\:Y_{j} \}\:\:\exp \{ (\imath/2)\:\sum_{h=(N-2)^{2}}^{(N-1)^{2}-1} z_{h}\:Z_{h} \} \ldots \nonumber \\ \ldots \exp \{ (\imath/2)\:\sum_{k=4}^{8} w_{k}\:W_{k} \}\:\:\exp \{ (\imath/2)\:\sum_{p=1}^{3} v_{i}\:V_{i} \}\ . \label{INDUCTFACTOR}
\end{eqnarray}

We can rewrite equation (\ref{INDUCTFACTOR}) using the induction process $S_{sub} = S_{N-1}\:\ldots \:S_{3}\:S_{2}$ as,

\begin{eqnarray}
\exp \{ (\imath/2)\:\sum_{i=1}^{N^{2}-1} x_{i}\:X_{i} \}  = S_{N}\:S_{N-1}\:\ldots \:S_{3}\:S_{2} \ , \label{FACTOR2}
\end{eqnarray}

where $S_{N}$ is the coset representative $\exp \{ (\imath/2)\:\sum_{j=(N-1)^{2}}^{N^{2}-1} y_{j}\:Y_{j} \}$ and by induction $S_{N-1}$ is the next coset representative $\exp \{ (\imath/2)\:\sum_{h=(N-2)^{2}}^{(N-1)^{2}-1} z_{h}\:Z_{h} \}$ in equation (\ref{INDUCTFACTOR}) and so on until we reach $S_{2}$ which is an actual element of the local group $SU(2)$. We parametrize the group $SU(N)$ as a sequence of local coset representative factors $S_{N}\:S_{N-1}\:\ldots \:S_{3}$ times the whole local group $SU(2)$. One local coset representative per coset in the factorization $S_{N}\:S_{N-1}\:\ldots \:S_{3}$ times the whole group $SU(2)$. An induction construction following the pattern for $SU(3)$ developed in reference \cite{ASU3}. We can then rewrite equation (\ref{INDUCTFACTOR}) or its compact form (\ref{FACTOR2}) as follows,

\begin{eqnarray}
\exp \{ (\imath/2)\:\sum_{i=1}^{N^{2}-1} x_{i}\:X_{i} \}  = S_{2}\:(S_{2}^{-1}\:S_{N}\:S_{2})\:(S_{2}^{-1}\:S_{N-1}\:S_{2})\:S_{2}^{-1}\ldots S_{2}\:(S_{2}^{-1}\:S_{3}\:S_{2}) \ . \label{FACTOR3}
\end{eqnarray}

We have already proved in section \ref{appendixII} that the conjugation of a local coset representative by a local subgroup element is another local coset representative. Since the $S_{2} = SU(2)$ is a local subgroup of all local groups $SU(n)$ with $N \geq n > 2$, then the conjugation of a $S_{n}$ coset representative by the local subgroup $SU(2)$ will be another local coset representative element in the coset $S_{n}$. Therefore we can rewrite equation (\ref{FACTOR3}) as

\begin{eqnarray}
\exp \{ (\imath/2)\:\sum_{i=1}^{N^{2}-1} x_{i}\:X_{i} \}  = S_{2}\:\overline{S}_{N}\:\overline{S}_{N-1}\:\ldots \:\overline{S}_{3} \ . \label{FACTOR4}
\end{eqnarray}

where $\overline{S}_{N}$ is a new local coset representative of the kind $\exp \{ (\imath/2)\:\sum_{j=(N-1)^{2}}^{N^{2}-1} \overline{y}_{j}\:Y_{j} \}$ for some new local scalars $\overline{y}_{j}\:\:j=(N-1)^{2} \ldots N^{2}-1$. Equation (\ref{FACTOR4}) would allow to write for two consecutive local $SU(N)$ gauge transformations $S$ and $\hat{S}$,

\begin{eqnarray}
\lefteqn{ Tr[\hat{S}\:S\:\Sigma^{\alpha\beta}\:S^{-1}\:\hat{S}^{-1}\:S_{\alpha}^{\:\:\rho}\: S_{\beta}^{\:\:\lambda}\:\ast \epsilon_{\rho\sigma}\:\ast \epsilon_{\lambda\tau}\:A^{\tau}] } \nonumber \\ && = Tr[(\hat{S}_{N}\:\hat{S}_{N-1}\:\ldots \:\hat{S}_{3}\:\hat{S}_{2})\:(S_{N}\:S_{N-1}\:\ldots \:S_{3}\:S_{2})\:\Sigma^{\alpha\beta}\:(S_{N}\:S_{N-1}\:\ldots \:S_{3}\:S_{2})^{-1}\:(\hat{S}_{N}\:\hat{S}_{N-1}\:\ldots \:\hat{S}_{3}\:\hat{S}_{2})^{-1}\: \nonumber \\ && S_{\alpha}^{\:\:\rho}\: S_{\beta}^{\:\:\lambda}\:\ast \epsilon_{\rho\sigma}\:\ast \epsilon_{\lambda\tau}\:A^{\tau}] = \nonumber \\ && Tr[(\hat{S}_{N}\:\hat{S}_{N-1}\:\ldots \:\hat{S}_{3})\:(\overline{S}_{N}\:\overline{S}_{N-1}\:\ldots \:\overline{S}_{3})\:(\hat{S}_{2}^{-1}\:S_{2})\:\Sigma^{\alpha\beta}\:(S_{2}^{-1}\:\hat{S}_{2})\:(\overline{S}_{N}\:\overline{S}_{N-1}\:\ldots \:\overline{S}_{3})^{-1}\:(\hat{S}_{N}\:\hat{S}_{N-1}\:\ldots \:\hat{S}_{3})^{-1}\: \nonumber \\ && S_{\alpha}^{\:\:\rho}\: S_{\beta}^{\:\:\lambda}\:\ast \epsilon_{\rho\sigma}\:\ast \epsilon_{\lambda\tau}\:A^{\tau}] =  Tr[\hat{\Lambda}^{\alpha}_{2\:\:\kappa}\:\hat{\Lambda}^{\beta}_{2\:\:\Omega}\:\tilde{\Lambda}^{\kappa}_{1\:\:\delta}\:
\tilde{\Lambda}^{\Omega}_{1\:\:\gamma}\:\nonumber \\ && (\hat{S}_{N}\:\hat{S}_{N-1}\:\ldots \:\hat{S}_{3})\:(\overline{S}_{N}\:\overline{S}_{N-1}\:\ldots \:\overline{S}_{3})\:\Sigma^{\alpha\beta}\:(\overline{S}_{N}\:\overline{S}_{N-1}\:\ldots \:\overline{S}_{3})^{-1}\:(\hat{S}_{N}\:\hat{S}_{N-1}\:\ldots \:\hat{S}_{3})^{-1}\: \nonumber \\ && S_{\alpha}^{\:\:\rho}\: S_{\beta}^{\:\:\lambda}\:\ast \epsilon_{\rho\sigma}\:\ast \epsilon_{\lambda\tau}\:A^{\tau}]\ . \label{FACTOR5}
\end{eqnarray}

The notation is $S_{2}\:\Sigma^{\alpha\beta}\:S_{2}^{-1} = \tilde{\Lambda}^{\alpha}_{1\:\:\delta}\:
\tilde{\Lambda}^{\beta}_{1\:\:\gamma}\:\Sigma^{\delta\gamma}$ and $\hat{S}_{2}^{-1}\:\Sigma^{\alpha\beta}\:\hat{S}_{2} = \hat{\Lambda}^{\alpha}_{2\:\:\delta}\:\hat{\Lambda}^{\beta}_{2\:\:\gamma}\:\Sigma^{\delta\gamma}$. For the sake of simplicity we are using the notation, $\Lambda^{(-1)\,\alpha}_{\:\:\:\:\:\:\:\:\:\:\:\:\delta} = \tilde{\Lambda}^{\alpha}_{\:\:\:\delta}$. Now, let us make use of the local transformation properties of the objects $\Sigma^{\alpha\beta}$, see section appendix II in reference \cite{A2}, and write for just a single local $SU(N)$ transformation $S = S_{N}\:S_{N-1}\:\ldots \:S_{3}\:S_{2} = S_{cf}\:S_{2}$, where for the sake of compactness we write $S_{cf} = S_{N}\:S_{N-1}\:\ldots \:S_{3}$ with $cf$ standing for coset factors.

\begin{eqnarray}
Tr[\tilde{\Lambda}^{\alpha}_{\:\:\:\delta}\:\tilde{\Lambda}^{\beta}_{\:\:\:\gamma}\:S_{cf}\:\Sigma^{\delta\gamma}\:S_{cf}^{-1}\:S_{\alpha}^{\:\:\rho}\: S_{\beta}^{\:\:\lambda}\:\ast \epsilon_{\rho\sigma}\:\ast \epsilon_{\lambda\tau}\:A^{\tau}] + \nonumber \\ {\imath \over g} \: Tr[\tilde{\Lambda}^{\alpha}_{\:\:\:\delta}\:\tilde{\Lambda}^{\beta}_{\:\:\:\gamma}\:S_{cf}\:\Sigma^{\delta\gamma}\:S_{cf}^{-1}\:S_{\alpha}^{\:\:\rho}\: S_{\beta}^{\:\:\lambda}\:\ast \epsilon_{\rho\sigma}\:\ast \epsilon_{\lambda\tau}\:\partial^{\tau}\:(S_{cf}\:S_{2})\:(S_{cf}\:S_{2})^{-1}] \ .
\label{S1L}
\end{eqnarray}

For the purpose of comparison in manuscript \cite{ASU3} we wrote $S_{2} = S_{ROT}$ since the local Lorentz transformations associated with local elements of the $SU(2)$ group are spatial rotations. Here since we are using induction over $SU(N)$ and not just $SU(3)$ we decided to generalize the notation. Making use now of the notation introduced in section \ref{QS} we can rewrite expression (\ref{S1L}) as,

\begin{eqnarray}
Tr[\tilde{\Lambda}^{\alpha}_{\:\:\:\delta}\:\tilde{\Lambda}^{\beta}_{\:\:\:\gamma}\:\Sigma_{2N-1}^{\delta\gamma}\:S_{\alpha}^{\:\:\rho}\: S_{\beta}^{\:\:\lambda}\:\ast \epsilon_{\rho\sigma}\:\ast \epsilon_{\lambda\tau}\:A^{\tau}] + \nonumber \\ {\imath \over g} \: Tr[\tilde{\Lambda}^{\alpha}_{\:\:\:\delta}\:\tilde{\Lambda}^{\beta}_{\:\:\:\gamma}\:\Sigma_{2N-1}^{\delta\gamma}\:S_{\alpha}^{\:\:\rho}\: S_{\beta}^{\:\:\lambda}\:\ast \epsilon_{\rho\sigma}\:\ast \epsilon_{\lambda\tau}\:\partial^{\tau}\:(S_{cf}\:S_{2})\:(S_{cf}\:S_{2})^{-1}] \ .
\label{S1L2}
\end{eqnarray}

The object $\Sigma_{2N-1}^{\delta\gamma}$ is a $\Sigma^{\delta\gamma}$ object transformed by a local coset representative $\Sigma_{2N-1}^{\delta\gamma} = S_{cf}\:\Sigma^{\delta\gamma}\:S_{cf}^{-1}$. The same object for all the local $SU(2)$ subalgebra. This locally gauge transformed gauge vector represented by equation (\ref{S1L2}) from a geometrical point of view has the following meaning \cite{A}$^{,}$\cite{A2}$^{,}$\cite{ASU3}. Let us focus on blade one, on blade two the analysis is analogous.

Our first conclusion from the results above, is that $SU(N)$ local gauge transformations, generate the composition of two transformations. First, there is a local tetrad transformation, generated by a locally inertial coordinate transformation $\tilde{\Lambda}^{\alpha}_{\:\:\:\delta}$, of the $SU(N-1)$ tetrads $S_{\alpha}^{\rho}$ inside the gauge vector. Second, the normalized tetrad vectors that generate blade one, which the vector (\ref{gaugev}) is gauging, undergo a LB1 transformation on the blade they generate because of the second line in equation (\ref{S1L2}). The two normalized vectors, that generate blade one, end up on the same blade one generated by the original normalized generators of the blade, $\left({Q_{(1)}^{\mu} \over \sqrt{-Q_{(1)}^{\nu}\:Q_{(1)\nu}}}, {Q_{(2)}^{\mu} \over \sqrt{Q_{(2)}^{\nu}\:Q_{(2)\nu}}}\right)$, after the LB1 local transformation (see the tetrad vectors introduced in (\ref{S1}-\ref{S4})). This second transformation is generated by the second line in (\ref{S1L2}). Therefore, the gauge invariance of the metric tensor is assured as was discussed in papers \cite{A}$^{,}$\cite{A2}$^{,}$\cite{ASU3}. We can continue making relevant remarks about these tetrad transformations. Within the set of LB1 tetrad transformations  of the pair $\left({Q_{(1)}^{\mu} \over \sqrt{-Q_{(1)}^{\nu}\:Q_{(1)\nu}}}, {Q_{(2)}^{\mu} \over \sqrt{Q_{(2)}^{\nu}\:Q_{(2)\nu}}}\right)$, there is an identity transformation that corresponds to the identity in $SU(N)$. To every LB1 tetrad transformation, which in turn is generated by $S_{cf}\:S_{2}$ in $SU(N)$, there corresponds an inverse, generated by $S_{2}^{-1}\:S_{cf}^{-1}$. We will prove in section \ref{appendixII} that the inverse $S_{2}^{-1}\:S_{cf}^{-1}$ can be reexpressed as a coset element times a subalgebra element, that is, can be rewritten in the same original coset parametrization. We observe also the following. Since locally inertial coordinate Lorentz transformations $\tilde{\Lambda}^{\alpha}_{\:\:\:\delta}$ of the $SU(N-1)$ tetrads $S_{\alpha}^{\rho}$ in general do not commute, then the locally $SU(N)$ tetrad generated transformations are non-Abelian. The non-Abelianity of $SU(N)$ is mirrored by the non-commutativity of these locally inertial coordinate transformations $\tilde{\Lambda}^{\alpha}_{\:\:\:\delta}$, which are essentially local space rotations. The key role in this non-commutativity is played by the object $\Sigma^{\alpha\beta}$, that translates the local $SU(2)$ subalgebra factor in local $SU(N)$ gauge transformations, into locally inertial Lorentz transformations. Another issue of relevance is related to the analysis of the ``memory'' of these transformations. As we did in papers \cite{A2}$^{,}$\cite{ASU3} we would like to know explicitly, if a second LB1 tetrad transformation, generated by a local gauge transformation $\hat{S}_{cf}\:\hat{S}_{2}$, will ``remember'' the existence of the first one, generated by $S_{cf}\:S_{2}$. To this end and following the lines in \cite{A2}, let us just write for instance the vector $Q_{(1)}^{\:\mu}$ after these two gauge transformations,

%\newpage

\begin{eqnarray}
\lefteqn{Q_{(1)}^{\:\mu} \rightarrow  \Omega^{\mu\nu}\:\Omega^{\sigma}_{\:\:\nu}\: Tr[\tilde{\Lambda}^{\alpha}_{2\:\:\kappa}\:\tilde{\Lambda}^{\beta}_{2\:\:\Omega}\:\tilde{\Lambda}^{\kappa}_{1\:\:\delta}\:
\tilde{\Lambda}^{\Omega}_{1\:\:\gamma}\:(S_{2N-1}\:\Sigma^{\delta\gamma}\:S_{2N-1}^{-1})\:S_{\alpha}^{\:\:\rho}\: S_{\beta}^{\:\:\lambda}\:\ast \epsilon_{\rho\sigma}\:\ast \epsilon_{\lambda\tau}\:A^{\tau}] +} \nonumber \\
&&{\imath \over g} \:\Omega^{\mu\nu}\:\Omega^{\sigma}_{\:\:\nu}\: Tr[\tilde{\Lambda}^{\alpha}_{2\:\:\kappa}\:\tilde{\Lambda}^{\beta}_{2\:\:\Omega}\:\tilde{\Lambda}^{\kappa}_{1\:\:\delta}\:
\tilde{\Lambda}^{\Omega}_{1\:\:\gamma}\:(S_{2N-1}\Sigma^{\delta\gamma}\:S_{2N-1}^{-1})\:S_{\alpha}^{\:\:\rho}\: S_{\beta}^{\:\:\lambda}\:\ast \epsilon_{\rho\sigma}\:\ast \epsilon_{\lambda\tau}\:\partial^{\tau}\:(S_{cf}\:S_{2})\:(S_{cf}\:S_{2})^{-1}] + \nonumber \\
&&{\imath \over g} \:\Omega^{\mu\nu}\:\Omega^{\sigma}_{\:\:\nu}\: Tr[\tilde{\Lambda}^{\alpha}_{2\:\:\delta}\:\tilde{\Lambda}^{\beta}_{2\:\:\gamma}\:(\hat{S}_{cf}\:\Sigma^{\delta\gamma}\:\hat{S}_{cf}^{-1})\:S_{\alpha}^{\:\:\rho}\:
S_{\beta}^{\:\:\lambda}\:\ast \epsilon_{\rho\sigma}\:\ast \epsilon_{\lambda\tau}\:\partial^{\tau}\:(\hat{S}_{cf}\:\hat{S}_{2})\:(\hat{S}_{cf}\:\hat{S}_{2})^{-1}] \ .
\label{S1DOUBLETR}
\end{eqnarray}

where $S_{2N-1} = S_{cf}\:S_{2}\:\hat{S}_{cf}\:S_{2}^{-1}$. It is clear that in the third line in (\ref{S1DOUBLETR}) neither $S_{cf}$ or $S_{2}$ are present. This third line represents the second LB1 transformation. In the second line through $S_{2N-1}$, only $\hat{S}_{cf}$ is present with regards to the second $\hat{S}_{cf}\:\hat{S}_{2}$ local $SU(N)$ gauge transformation. Now, the point is that $\hat{S}_{cf}$ is the coset representative for the second $SU(N)$ local gauge transformation. It is given by a sequence of vectors in the $S^{2N-1}\ldots S^{5}$ spheres, the quotient space, see sections \ref{QS}, \ref{appendixI} and \ref{appendixII}.

%It is also clear that $S_{5}$ is by itself a coset element in the quotient space given that we can write

%\begin{equation}
%(S_{1}\:S_{ROT1})\:(S_{2}\:S_{ROT2}) = S_{5}\:S_{ROT1}\:S_{ROT2}. \label{cosetprod}
%\end{equation}

Therefore, the second subgroup spanned by the elements $\hat{S}_{2}$, the local $SU(2)$ subgroup to $SU(N)$, is not present in the second line that represents the first LB1 local transformation. $\hat{S}_{cf}$ is one coset element representing a fixed direction in the local sequence of n-spheres $S^{2n-1}, n=3 \ldots N$, just one for the whole of the local $SU(2)$ subgroup spanned by all the $\hat{S}_{2}$. Therefore the second local $SU(N)$ gauge transformation is only present through the local equivalence class coset representative. With regard to the object $S_{2}\:\hat{S}_{cf}\:S_{2}^{-1}$, we will study its properties in section \ref{appendixII}. We will prove that this object is a local coset representative by itself. Therefore $S_{2N-1} = S_{cf}\:S_{2}\:\hat{S}_{cf}\:S_{2}^{-1}$ is nothing but the product of two coset representatives.

We can notice that the second line contains the same $SU(2)$ Lorentz transformed tetrads $S_{\alpha}^{\:\:\rho}$ as the first one. Therefore, when we compare these two terms, it is straightforward to see that it is not possible, after the second gauge transformation, from these transformed $SU(2)$ tetrads, to ``remember'' any relative change associated to the second gauge transformation.

In addition, the second line in (\ref{S1DOUBLETR}) contains in the derivative only $S_{cf}\:S_{2}$, and the third line contains only $\hat{S}_{cf}\:\hat{S}_{2}$. This means that the second LB1 tetrad transformation on blade one will not remember the first one. The algebra underlying these statements can be followed through \cite{A}. Again, as reasoned in \cite{A2}$^{,}$\cite{ASU3}, another way of thinking of (\ref{S1DOUBLETR}) is by first performing two successive local Lorentz transformations $\tilde{\Lambda}^{\alpha}_{2\:\:\kappa}\:\tilde{\Lambda}^{\kappa}_{1\:\:\delta}$ of the $SU(N-1)$ tetrad $S_{\alpha}^{\:\:\rho}$ and similar for the tetrad $S_{\beta}^{\:\:\lambda}$ in the first line, and second, by performing two successive $LB1$ tetrad transformations in the second and third line.

We could repeat exactly the remainder of the analysis done in papers \cite{A2}$^{,}$\cite{ASU3} in the section ``gauge geometry'' and reach similar conclusions, the only specific and distinctive issue of surjectivity in the local $SU(N)$ case that should be highlighted and discussed carefully has already been studied in detail in the section VIII of reference \cite{ASU3}. Finally we are able to state two new theorems.

\newtheorem {guesslb1} {Theorem}
\newtheorem {guesslb2}[guesslb1] {Theorem}
\begin{guesslb1}
The mapping between the local gauge group of transformations $SU(N)$ and the tensor product of the $N^{2}-1$ local groups of LB1 transformations is isomorphic. \end{guesslb1}

Following analogously the reasoning laid out in references \cite{A}$^{,}$\cite{A2}$^{,}$\cite{ASU3}, in addition to the ideas above, we can also state,

\begin{guesslb2}
The mapping between the local gauge group of transformations $SU(N)$ and the tensor product of the $N^{2}-1$ local groups of LB2 transformations is isomorphic. \end{guesslb2}

%The group $SU(N)$ is connected to the identity making unnecessary to discuss isomorphisms of group sheet components as in the $SU(2)$ case.
It is very important that we make a clarification on an issue that might give rise to confusion.
%It is clear that if we have a local plane one or two spanned by tetrad vectors with just a single and only extremal field-gauge vector structure and we consider only on this plane one-dimensional groups of transformations, such as LB1 for the local plane one or LB2 for the local plane two, which is nothing but $SO(2)$, and furthermore, we consider tensor products of these tetrad local groups of transformations on this particular local plane, we can conclude that these tensor products are reduced to simple LB1 or LB2 local groups of transformations. This is not what we mean in the enunciation of theorems (1-2).
In theorems (1-2) we are considering $N^{2}-1$ copies of the same spacetime, and a different tetrad at the same point in each spacetime copy. These $N^{2}-1$ tetrads have a similar extremal field-gauge vector structure as in (\ref{S1}-\ref{S4}). They are normalized and a choice of gauge vector has been made. But they are not the same. They could be Lorentz transformed into each other, under non-trivial Lorentz spatial rotations, for instance. We know from manuscript \cite{A2} that a local Lorentz transformation of a tetrad with an extremal field-gauge vector structure transforms into another tetrad with a similar extremal field-gauge vector structure. Therefore, we are considering $N^{2}-1$ local tetrads at the same spacetime point which are not the same for different copies. The local planes one and two will be tilted with respect to each other. This is what we mean by $N^{2}-1$ LB1 or LB2 groups under tensor product. Now, from a practical point of view what we also mean by these theorems is that we are able to reconstruct the original local $SU(N)$ gauge transformation by knowing for example, the boosts in the LB1 case, or the spatial rotations in the LB2 case, for the tetrad local transformations at the point under consideration. By knowing the local Lorentz transformation values for the $N^{2}-1$ copies, and given all the fields, specially the $N^{2}-1$ tetrads at the same point, we can reconstruct the local $SU(N)$ transformation that gave rise either to $N^{2}-1$ local LB1 transformations or independently to $N^{2}-1$ local LB2 spatial transformations. Because the theorems proved, represent local isomorphisms between either $N^{2}-1$ LB1 groups and $SU(N)$ or independently $N^{2}-1$ LB2 groups and $SU(N)$.

\section{Conclusions}
\label{conclusions}

The new tetrads introduced in manuscript \cite{A} with the purpose of simplification for the Einstein-Maxwell geometries have played a dual role. On the one hand these tetrads maximally simplify the expressions for the electromagnetic, stress-energy and metric tensors, the objects that carry the physical information. On the other hand these new tetrads enable the proofs for new fundamental group isomorphisms and homomorphisms. These new isomorphisms and homomorphisms are directly linked to the internal symmetries of the Standard Model in the Abelian and also extended to the non-Abelian cases \cite{A,A2,A3,ASU3}. These new tetrads carry gravitational information on one hand and on the other hand display the geometrical nature of the local groups of internal transformations of the Standard Model. It is noticed from the outset that since these tetrads enable the realization in four-dimensional Lorentzian and curved spacetimes of both the local groups of spacetime symmetries and also the local groups of gauge symmetries that the assumptions and hypotheses made at the outset of the no-go theorems \cite{SWNG,LORNG,CMNG} of the sixties are incorrect. Therefore, the no-go theorems are incorrect, as we have already said in reference \cite{IWCP} not because of the internal logic of these theorems but because of the hypotheses made at the beginning of their proofs. With all this material in mind we set out to further prove that the local groups of flavor transformations for $N > 3$ are isomorphic or homomorphic to the local groups of these new tetrad transformations on both the locals plane one and the orthogonal plane two. This proof by transitivity of the isomorphic relation between the local spacetime groups of tetrad transformations LB1 and LB2 or their tensor products for the connected sheet or homomorphic relation for the whole of LB1 or its tensor products and LB2 or their tensor products, set the stage for grand field unification not only for the Standard Model $SU(3) \times SU(2) \times U(1)$ case, but also for the extension to $SU(N)$ in general or $SU(N)\times \ldots \times SU(3) \times SU(2) \times U(1)$ as well. Incorporating all these internal local groups of transformations into the Riemannian geometry in a natural and non-trivial way and also incorporating the gauge theories into the Riemannian geometry also in a natural way. At the base of all these constructions and analysis are the two elements that intervene in the construction of these new tetrads. The structure of these new tetrads is made of the tetrad skeletons and the gauge vector fields, see section \ref{intro}. The skeletons are local gauge invariants while the gauge vectors are gauge by themselves and contain gauge in their construction. It is through these two elements that enable the construction of these new tetrads that we can simplify the tensor expressions and establish many proofs about local internal and spacetime groups of transformations. We achieve unification in four-dimensional curved Lorentz spacetimes through the use of these new objects carriers of all physical information regarding Standard Model gauge fields and also gravitational fields, simultaneously. It is not only grand field unification, it is also grand group unification. It is in this way that we managed to link or couple in a single structure the fields describing long range and short range interactions. In fact we have implicitly proved that we can associate spacetimes to microparticles since these spacetimes depict manifestly all the local gauge symmetries of the Standard Model. Microparticles in the end have associated gravitational fields. Now, the question stands about how these classical results can match the well known quantum perturbative phenomena. The answer is in manuscripts \cite{dsmg,dsmg1} were it is proved that interactions between agents that curve spacetime destroy local symmetries by perturbative action through the local tilting of the two orthogonal planes of symmetry. The good news is that the new local and tilted planes of symmetry describe new local analogous symmetries. Symmetries are instantaneously destroyed in interactions that simultaneously create new ones. There is a symmetry evolution through perturbative interactions. These perturbative interactions are generic and they could be quantum fluctuations continuous or discrete since the theorems that prove that there is a local symmetry evolution consider generic local perturbations. Therefore we have achieved the unification with the quantum realm by considering generic perturbations and analyzing the local geometry during the perturbative process.

\section{Appendix I}
\label{appendixI}

We will discuss in this first appendix the parametrization of the local $SU(N)$ elements that represent the quotient $SU(N) / SU(N-1)$. We start by building the local $SU(N)$ group element objects $S_{N} = \exp \{ (\imath/2)\:\sum_{i=(N-1)^{2}}^{N^{2}-1} \theta_{i}\:X_{i} \}$. The $\theta_{i}, i=(N-1)^{2} \cdots N^{2}-1$ are all local scalars. The $X_{i}, i=(N-1)^{2} \cdots N^{2}-1$ are the remaining $2N-1$ $SU(N)$ group generators, other than the $(N-1)^{2}-1$ in the local $SU(N-1)$ subalgebra. If the exponential matrix is fully calculated, from the condition that $\det S_{N}=1$ it emerges that there is an isomorphism between $SU(N) / SU(N-1)$ and $S^{2N-1}$. We will take advantage of this isomorphism to reparameterize the original quotient elements $S_{N}$. Let us introduce the coordinates of the stereographic projections for the unit sphere $S^{2N-1}$. The local ($2N-1$)-sphere is defined through $\sum_{i=1}^{2N} x_{i}^{2} = 1$ where the $x_{i},\: i=1 \cdots 2N$ are local coordinates. Following closely chapter III in \cite{CBDW} an in order to construct an atlas we let $P$ and $Q$ be the north and south poles respectively. Let the point $P=(0,\ldots,1)$ and $Q=(0,\ldots,-1)$. Let $U=S^{2N-1}-{P}$ and $V=S^{2N-1}-{Q}$, let g and h be the stereographic projections of the poles $P$ and $Q$ on the plane $x_{2N}=0$,

\begin{center}
$\: g: U \rightarrow \Re^{2N-1} \:\:\: by  \:\:\:  y_{i} = x_{i} / (1-x_{2N}) \:\:\: for  \:\:\:  i=1 \cdots 2N-1$
\end{center}

\begin{center}
$\: h: V \rightarrow \Re^{2N-1} \:\:\: by  \:\:\: z_{i} = x_{i} / (1+x_{2N}) \:\:\: for  \:\:\: i=1 \cdots 2N-1$
\end{center}

See \cite{CBDW} for the proof that it is an atlas. Both g and h have range $\Re^{2N-1}$. We can think $S^{2N-1}$ as $\Re^{2N-1}$ plus a single point at infinity. Then, knowing that the coordinates are local because they are defined at every point in spacetime, we can replace the local scalars $\theta_{i}, i=(N-1)^{2} \cdots N^{2}-1$ for the local atlas coordinates. We might ask what is the advantage of this procedure that feedsback new parameters for the local coset elements. We believe that the local coset representatives parameterized in terms of the local stereographic projections present a clear relationship between the coset elements and the local unit sphere $S^{2N-1}$. It is also evident that when we calculate fully the matrix for $S_{N} = \exp \{ (\imath/2)\:\sum_{i=(N-1)^{2}}^{N^{2}-1} y_{i}\:X_{i} \}$ or $S_{N} = \exp \{ (\imath/2)\:\sum_{i=(N-1)^{2}}^{N^{2}-1} z_{i}\:X_{i} \}$ we will get again for the condition $\det S_{N}=1$ a new equation that signals the isomorphism to $S^{2N-1}$ but now in new coordinates. We simply reparameterized the local coset elements in terms of the local stereographic projections atlas coordinates in order to establish a clear relationship between the quotient space $SU(N) / SU(N-1)$, that is the unit ($2N-1$)-sphere, and the coset representatives. For a further advantage of this coset parametrization see section \ref{appendixII}.

\section{Appendix II}
\label{appendixII}

We will show in this section that the result obtained in section VII of manuscript \cite{ASU3} for SU(3) can be generalized to $SU(N)$. That is, if the $SU(N)$ group elements can be parametrized as,

\begin{eqnarray}
\exp \left\{Y\right\}\:\exp \left\{X\right\} = \exp \{ (\imath/2)\:\sum_{j=(N-1)^{2}}^{N^{2}-1} \Psi_{j}\:X_{j} \}\:\:\exp \{ (\imath/2)\:\sum_{i=1}^{(N-1)^{2}-1} \theta_{i}\:X_{i} \} \ . \label{PARAM}
\end{eqnarray}

The  $\theta_{i}, i=1 \cdots (N-1)^{2}-1$ and $(2N-1)$ $\Psi_{j}, j=(N-1)^{2}\cdots N^{2}-1$ are all local scalars. The first object $\exp\left\{X\right\}$ belongs to the local subgroup $SU(N-1)$ and the second one $\exp\left\{Y\right\}$ is a local coset representative. We name the generators following the same order as in reference \cite{GMS}. Then, for example, $X_{8}$ is the diagonal one $\left(\frac{1}{\sqrt{3}}, \frac{1}{\sqrt{3}}, \frac{-2}{\sqrt{3}}, 0\ldots0 \right)$ with the last zero in the Nth place. We would like to study the object $\exp{\left\{X\right\}}\:\exp{\left\{Y\right\}}\:\exp{\left\{-X\right\}}$ since this result would be useful at several stages in our analysis. To this end, first we will analyze all possible commutators between the generators of $SU(N)$ that do not belong to $SU(N-1)$ and the generators that belong to $SU(N-1)$ following all the conventions and notation of reference \cite{GMS} chapter XI. Let us remember that we are using this factorization (\ref{PARAM}) as a natural result of Cartan's method based on factor spaces like

\begin{equation}
SU(N) / SU(N-1) \cong S^{2N-1}\ , \label{quotiantN}
\end{equation}

with $N^{2}-1-((N-1)^{2}-1)=2N-1$ the number of generators in the second factor in the (\ref{PARAM}) parametrization. The parametrization of the second factor will follow the same prescription as in the case $SU(3) / SU(2) \cong S^{5}$ in manuscript \cite{ASU3} since we use the parametrization for the $S^{2N-1}$ sphere as the natural parametrization for the coset representatives given that the factor space in the $SU(N) / SU(N-1) \cong S^{2N-1}$ is precisely the $2N-1$ sphere. This present paper requires that we check as we did in paper \cite{ASU3} that the commutator of a generator in $SU(N-1)$ with the generators in $SU(N)$ outside $SU(N-1)$ gives a generator outside $SU(N-1)$. This result will be relevant in order to transform right cosets into left cosets given the fact that $SU(N-1)$ is not a normal subgroup of $SU(N)$. In turn these results will be useful in our induction proofs in the paper. We proceed then to introduce the generators of $SU(N)$,

\begin{eqnarray}
\left[{\widehat{\lambda}}^{(1)}(i,j)\right]_{\mu\nu} &=& \delta_{j\mu}\:\delta_{i\nu} + \delta_{j\nu}\:\delta_{i\mu} \label{generator1} \\
\left[{\widehat{\lambda}}^{(2)}(i,j)\right]_{\mu\nu} &=& \imath\:(\delta_{j\mu}\:\delta_{i\nu} - \delta_{j\nu}\:\delta_{i\mu}) \label{generator2} \\
\left[{\widehat{\lambda}}_{n^{2}-1}(i,j)\right]_{\mu\nu} &=& \sqrt{2 \over n^{2}-n}\:\left[\delta_{\mu\nu}\:step(\mu,n) - (n-1)\:\delta_{\mu\nu}\:\delta_{\nu n}\right] \ . \label{generator3}
\end{eqnarray}

where in equation (\ref{generator3}) $n = 2,3,\ldots N$ and $step(i,j) = 1 \:\: if \:\: i < j$ and $step(i,j) = 0 \:\: if \:\: i \geq j$. We define generators (\ref{generator1}-\ref{generator2}) for every $i,j = 1\ldots N, i < j$. The generators (\ref{generator3}) are not zero only for $i = j = n$ and for $n = 2,3,\ldots N$. The number of generators (\ref{generator1}) is $N\:(N-1)/2$. The number of generators (\ref{generator2}) is $N\:(N-1)/2$ as well. The number of generators (\ref{generator3}) is $N-1$ giving a total $N^{2}-1$. These matrices defined this way are a basis of the vector space of the traceless $N \times N$ matrices and make up a representation of the $SU(N)$ generators. In all matrices (\ref{generator1}-\ref{generator3}) the indices $\mu$ and $\lambda$ run from 1 to $N$. In all commutators below $\nu$ is a summing index from 1 to $N$. For $1 \leq i, j \leq N-1$ and $1 \leq k \leq N-1$ after some algebra we get the following results,

\begin{eqnarray}
\left[{\widehat{\lambda}}^{(1)}(i,j)\right]_{\mu\nu}\:\left[{\widehat{\lambda}}^{(1)}(k,N)\right]_{\nu\lambda} - \left[{\widehat{\lambda}}^{(1)}(k,N)\right]_{\mu\nu}\:\left[{\widehat{\lambda}}^{(1)}(i,j)\right]_{\nu\lambda} = \nonumber \\ = \imath\: \delta_{ik}\:\left[{\widehat{\lambda}}^{(2)}(j,N)\right]_{\mu\lambda} + \imath\:\delta_{jk}\:\left[{\widehat{\lambda}}^{(2)}(i,N)\right]_{\mu\lambda} \ . \label{comm11}
\end{eqnarray}

\begin{eqnarray}
\left[{\widehat{\lambda}}^{(1)}(i,j)\right]_{\mu\nu}\:\left[{\widehat{\lambda}}^{(2)}(k,N)\right]_{\nu\lambda} - \left[{\widehat{\lambda}}^{(2)}(k,N)\right]_{\mu\nu}\:\left[{\widehat{\lambda}}^{(1)}(i,j)\right]_{\nu\lambda} = \nonumber \\ = -\imath\: \delta_{ik}\:\left[{\widehat{\lambda}}^{(1)}(j,N)\right]_{\mu\lambda} - \imath\:\delta_{jk}\:\left[{\widehat{\lambda}}^{(1)}(i,N)\right]_{\mu\lambda} \ . \label{comm12}
\end{eqnarray}

\begin{eqnarray}
\left[{\widehat{\lambda}}^{(2)}(i,j)\right]_{\mu\nu}\:\left[{\widehat{\lambda}}^{(2)}(k,N)\right]_{\nu\lambda} - \left[{\widehat{\lambda}}^{(2)}(k,N)\right]_{\mu\nu}\:\left[{\widehat{\lambda}}^{(2)}(i,j)\right]_{\nu\lambda} = \nonumber \\ = \imath\: \delta_{ik}\:\left[{\widehat{\lambda}}^{(2)}(j,N)\right]_{\mu\lambda} - \imath\:\delta_{jk}\:\left[{\widehat{\lambda}}^{(2)}(i,N)\right]_{\mu\lambda} \ . \label{comm22}
\end{eqnarray}

The two next commutators will be considered for general $1 \leq p,q \leq N$ even though the cases that interest us will be for $1 \leq p \leq N-1$ and $q = N$.

\begin{eqnarray}
\left[{\widehat{\lambda}}_{n^{2}-1}(n,n)\right]_{\mu\nu}\:\left[{\widehat{\lambda}}^{(1)}(p,q)\right]_{\nu\lambda} - \left[{\widehat{\lambda}}^{(1)}(p,q)\right]_{\mu\nu}\:\left[{\widehat{\lambda}}_{n^{2}-1}(n,n)\right]_{\nu\lambda} = \nonumber \\ = -\imath\:\sqrt{2 \over n^{2}-n}\:\{step(q,n)-step(p,n)-(n-1)\:[\delta_{qn}-\delta_{pn}]\}\:\left[{\widehat{\lambda}}^{(2)}(p,q)\right]_{\mu\lambda} \ . \label{comm1n}
\end{eqnarray}

\begin{eqnarray}
\left[{\widehat{\lambda}}_{n^{2}-1}(n,n)\right]_{\mu\nu}\:\left[{\widehat{\lambda}}^{(2)}(p,q)\right]_{\nu\lambda} - \left[{\widehat{\lambda}}^{(2)}(p,q)\right]_{\mu\nu}\:\left[{\widehat{\lambda}}_{n^{2}-1}(n,n)\right]_{\nu\lambda} = \nonumber \\ = \imath\:\sqrt{2 \over n^{2}-n}\:\{step(q,n)-step(p,n)-(n-1)\:[\delta_{qn}-\delta_{pn}]\}\:\left[{\widehat{\lambda}}^{(1)}(p,q)\right]_{\mu\lambda} \ . \label{comm2n}
\end{eqnarray}

As we can see in equations (\ref{comm11}-\ref{comm2n}), the commutators of any generator in $SU(N-1)$ with the generators of any generators in $SU(N)$ that do not belong to $SU(N-1)$ will result in a generator outside $SU(N-1)$. That is, all the commutators of any generator in $\exp \left\{X\right\} = \exp \{ (\imath/2)\:\sum_{i=1}^{(N-1)^{2}-1} \theta_{i}\:X_{i} \}$ with any generator in $\exp \left\{Y\right\} = \exp \{ (\imath/2)\:\sum_{j=(N-1)^{2}}^{N^{2}-1} \Psi_{j}\:X_{j} \}$ gives a generator in $\exp \left\{Y\right\}$.  The first object $\exp\left\{X\right\}$ belongs to the local subgroup and the second one $\exp\left\{Y\right\}$ is a local coset representative. We name the generators following the same order as in reference \cite{GMS}. We would like to study the object $\exp{\left\{X\right\}}\:\exp{\left\{Y\right\}}\:\exp{\left\{-X\right\}}$. First we expand the exponential $\exp{\left\{Y\right\}} = \textbf{1} + Y + \frac{1}{2!}\:Y^{2} + \cdots$. Then we observe the following,

\begin{eqnarray}
\exp{\left\{X\right\}}\:\exp{\left\{Y\right\}}\:\exp{\left\{-X\right\}} = \exp{\left\{X\right\}}\:\left( \sum_{n=0}^{\infty} \frac{1}{n!}\:Y^{n} \right)\:\exp{\left\{-X\right\}} \nonumber \\
= \sum_{n=0}^{\infty} \frac{1}{n!}\:\exp{\left\{X\right\}}\:\left(Y^{n}\right)\:\exp{\left\{-X\right\}} =  \sum_{n=0}^{\infty} \frac{1}{n!}\:\left(\exp{\left\{X\right\}}\:Y\:\exp{\left\{-X\right\}}\right)^{n} \nonumber \\
= \exp \left( \exp{\left\{X\right\}}\:Y\:\exp{\left\{-X\right\}}\right) .\label{prevhad}
\end{eqnarray}

Now we proceed to apply the Hadamard formula \cite{RGLG} to the exponent in (\ref{prevhad}),

\begin{eqnarray}
\exp{\left\{X\right\}}\:Y\:\exp{\left\{-X\right\}} = Y + \left[X , Y\right] + \frac{1}{2!}\:\left[X , \left[X , Y\right]\right] + \cdots \label{hadamard}
\end{eqnarray}

Next we evaluate the commutator $\left[X , Y\right]$. If use the results (\ref{comm11}-\ref{comm2n}) we see very readily that $\left[X , Y\right]$ is an object of the $Y$ kind. Again through the use of equations (\ref{comm11}-\ref{comm2n}) the next commutator $\left[ X , \left[X , Y\right]\right]$ in the expansion (\ref{prevhad}) is another object of the $Y$ kind. Because of the commutators (\ref{comm11}-\ref{comm2n}) the Hadamard formula is telling us that the conjugation of a coset representative $\exp{\left\{Y\right\}}$ by a subgroup element $\exp{\left\{X\right\}}$ in $SU(N-1)$ is yielding another coset representative, which is exactly what we wanted to prove in this section. Therefore, when we sum the Hadamard expansion we obtain,

\begin{eqnarray}
\exp{\left\{X\right\}}\:Y\:\exp{\left\{-X\right\}} = (\imath/2)\:\sum_{p=(N-1)^{2}}^{N^{2}-1}\:\phi_{p}\:X_{p} \ ,  \label{sumhadamard}
\end{eqnarray}

for some $\phi_{p},\: p:(N-1)^{2}\ldots N^{2}-1$. Now, we can call,

\begin{eqnarray}
A = \sum_{p=(N-1)^{2}}^{N^{2}-1}\:\phi_{p}^{2} = \frac{\left(1 - \phi_{2N}^{2}\right)}{\left(1 \mp \phi_{2N}\right)^{2} }  \ .  \label{sixthcoord}
\end{eqnarray}

where the $\mp$ refers to the two posible charts in our parametrization, see section \ref{appendixI}. Finally we find,

\begin{eqnarray}
\phi_{2N} = \frac{\pm A \mp 1 }{ A + 1}  \ .  \label{sixthcoordfinal}
\end{eqnarray}

We conclude that the original object $\exp{\left\{X\right\}}\:\exp{\left\{Y\right\}}\:\exp{\left\{-X\right\}}$ is a coset representative. When we analyze the memory of the transformation in equation (\ref{S1DOUBLETR}) we can see that the object $S_{2}\:\hat{S}_{cf}\:S_{2}^{-1}$ is a coset element as well. Therefore, there is no memory in the subsequent transformations.

%\bibliography{your-bib-file} % place the references here.

\end{document}